\documentclass[ prl,
 amsmath,amssymb,
 reprint,%
]{revtex4-1}

\usepackage{graphicx}

\usepackage{float}

\begin{document}


\title{Nanophononic metamaterial: Thermal conductivity reduction by local resonance} 



\author{Bruce L. Davis}
\author{Mahmoud I. Hussein}
\email[]{mih@colorado.edu}
\affiliation{Department of Aerospace Engineering Sciences, University of Colorado Boulder, Colorado 80309, USA }

\date{\today}

\begin{abstract}
We present the concept of a locally resonant nanophononic metamaterial for thermoelectric energy conversion. Our configuration, which is based on a silicon thin-film with a periodic array of pillars erected on one or two of the free surfaces, qualitatively alters the base thin-film phonon spectrum due to a hybridization mechanism between the pillar local resonances and the underlying atomic lattice dispersion. Using an experimentally-fitted lattice-dynamics-based model, we conservatively predict a drop in the metamaterial thermal conductivity to as low as $50\%$ of the corresponding uniform thin-film value despite the fact that the pillars add more phonon modes to the spectrum.  
\end{abstract}


\maketitle 


The utilization of nanostructured materials for control of heat transport is a rapidly growing area of research~\cite{Balandin_1998,*Chen_2000}. Specifically, the manipulation of heat carrying phonons, or elastic waves that propagate and scatter at the nanoscale, can yield beneficial thermal properties. One pivotal application relates to thermoelectric materials, or the concept of converting energy in the form of heat into electricity and vice-versa. The ability to use nanostructuring to reduce the thermal conductivity, $k$, without negatively impacting the power factor, $S^2\sigma$  (where $S$ is the Seebeck coefficient and $\sigma$ is the electrical conductivity) provides a promising avenue for achieving high values of the thermoelectric energy conversion $ZT$ figure-of-merit~\cite{Dresselhaus_2007,*Snyder_2008}.\\
\indent The manipulation of elastic waves in a periodic medium can be realized primarily in two distinct ways: (1) the utilization of Bragg-scattering phononic crystals and (2) the introduction of local resonance. The latter, which is proposed here for the first time for the reduction of thermal conductivity, renders the medium a ``metamaterial". The concept of a phononic crystal~\cite{sigalas1993band,*kushwaha1993acoustic} involves a material with an artificial periodic internal structure for which the lattice spacing has a length scale on the order of the propagating waves, and hence wave interferences occur across the unit cell thus providing a unique frequency band structure with the possibility of band gaps. Focusing on nanoscale phonon transport, the periodic material can be realized in a variety of ways such as by the layering of multiple constituents, also known as a layered superlattice~\cite{Cleland_2001,*McGaughey_2006}, or the introduction of inclusions and/or holes, as in a nanophononic crystal (NPC)~\cite{Gillet_2009,*Tang_2010,*Yu_2010,*Hopkins_2011,*He_2011,*Davis_2011,*Robillard_2011}. The concept of a metamaterial, on the other hand, generally involves the inclusion of local resonators which enable unique subwavelength properties to emerge. While periodicity is advantageous, it is not necessary in a metamaterial. At the macroscale (where the focus is on acoustics or mechanical vibrations), locally resonant periodic metamaterials have been considered in various forms, such as by having heavy inclusions coated with a compliant material (e.g., rubber-coated lead spheres) hosted in a relatively lighter and less stiff matrix (e.g., epoxy)~\cite{liu2000locally}, or by the presence of pillars on a plate~\cite{pennec2008low,*wu2008evidence}.\\  
\indent In this Letter, we introduce the concept of a phononic metamaterial at the nanoscale, which we refer to as a nanophononic metamaterial (NPM). The goal is to significantly reduce the thermal conductivity in a nanostructured semiconducting material and to do so without affecting other important factors, especially the electrical conductivity. For both functional and practical purposes, we choose silicon thin-films as the foundation material for the creation of a locally resonant NPM. Using a reduced-dimension material such as a thin-film already causes a reduction of $k$ without necessarily impacting $S^2\sigma$~\cite{Liao_1999,*Venkatasubramanian_2001,*Hochbaum_2008,*Boukai_2008}, and is also favorable from the point of view of device integration. The choice of silicon is beneficial due to its wide use in the electronics industry and ease of fabrication; however other materials may be considered in the future. The resonators take the form of a periodic array of nanoscale pillars that extrude off the surface of the thin-film (on either one side or both sides, as practically permitted). Such structure has been considered in the literature for other applications and shown to be feasible from the point of view of fabrication~\cite{Chekurov_NanoTech_2009,*Henry_NanoTech_2010}. The primary advantage of this configuration is that the pillars exhibit numerous local resonances that couple, or more specifically hybridize, with the underlying atomic-level phonon dispersion of the thin-film and does so across the full range of its spectrum. These couplings drastically lower the group velocities and hence the thermal conductivity. This phenomenon is also known as avoided-crossing, which has been studied in natually occuring materials that has guest atoms encampsulated in caged structures such as clathrates~\cite{Dong_PRL_2001,*Christensen_NM_2008}. In contrast to a NPM, however, the hybridizations in these systems are limited to the modes of the guest atom and typically take place only within the acoustic range of the spectrum. Another important benefit to utilizing pillars is that the feature manipulating the coherent component of the thermal transport (i.e., the pillar itself) is physically outside of the primary flow path of the electrons (which happens in the main body of the thin film). This is a key advantage compared to thin-film-based NPCs, in which the inclusions or the holes penetrate through the thickness of the thin-film and hence may, in addition to scattering phonons, undesirably cause an obstruction to the electron transport. Thus, with the proposed pillared NPM configuration, the concern about the competition between coherent and incoherent thermal transport and how to elucidate the interplay of these two mechanisms (in order to enable a most effective thermoelectric material design) are no longer of critical importance. Yet, for thin-film-based NPCs, or other types of thermoelectric thin-flms with proven improvements in $ZT$ values such as superlattices of different forms~\cite{Dresselhaus_2007,*Snyder_2008}, the addition of a pillared array may be utilized as an ``over and above" augmentation to existing structures to provide a transformative improvement in performance.\\
\indent We begin our investigation with the creation of an atomic-level unit cell model for a uniform thin-film with thickness, $t$. We resort to a conventional cell (CC) description which consists of eight atoms packaged as a cube with side length $a = 0.54$ nm. Due to this conveniently shaped box-like structure, the CC will be used as our building block which we will replicate along an orthogonal simple cubic lattice to generate a supercell for the thin-film structure. This is done for the uniform thin-film and will be built upon later when a pillar is added to the free surface(s). For the uniform thin-film, the supercell consists of a vertical strip constructed by stacking $M$ CCs on top of each other along the out-of-plane $z$-direction. The dimensions of this supercell will be denoted by $A_{x}\times A_{y}\times A_{z}$, where $A_{x} = A_{y} = a$ and $A_{z} = Ma = t$. \\
\begin{figure}[b!]
\centering
\includegraphics[scale=1]{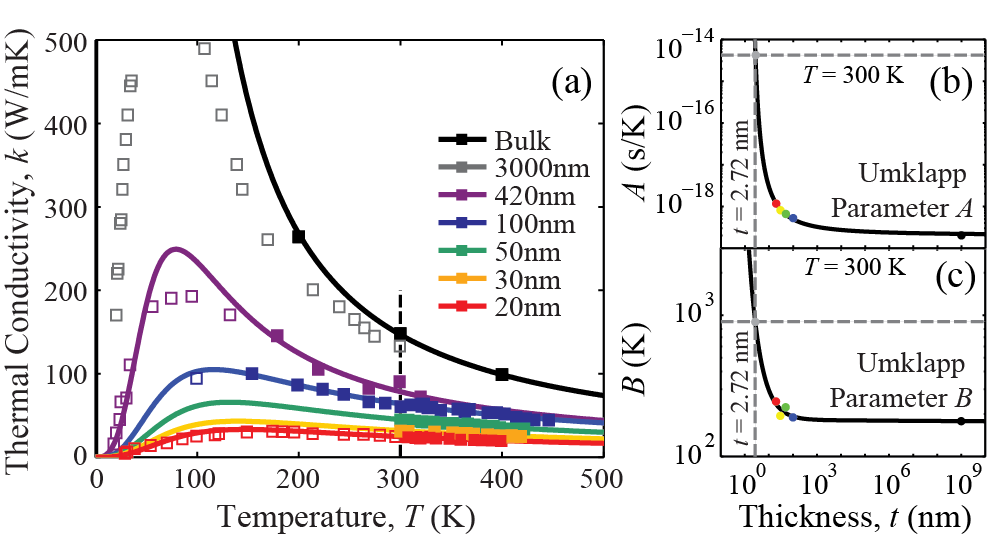}
\caption{(a) Thermal conductivity as a function of $T$ for various thin-films. The squares are measurements~\cite{Goodson_1999,*Liu_2004,*Liu_2005} and the solid lines represent the fitted model. (b) and (c) Umklapp scattering parameters, $A$ and $B$, respectively, as a function of $t$ at room temperature. These parameters are determined by fitting the thermal conductivity predictions with the empirical data points shown in (a). The C-H 2D thermal conductivity formulation, reinforced with full dispersion information, is shown to represent the experimental values very well for various $t$ values. For thin-films with thickness smaller than that is empirically available, an extrapolation is carried out. This is demonstrated in (b) and (c) for the thickness value of $t = 2.72$ nm, where $A = 4.14\times 10^{-15}$ s/K and $B = 899$ K.}
\label{fig:01}
\end{figure}
\begin{figure*}[ht]
\centering
\includegraphics[scale= 1.0]{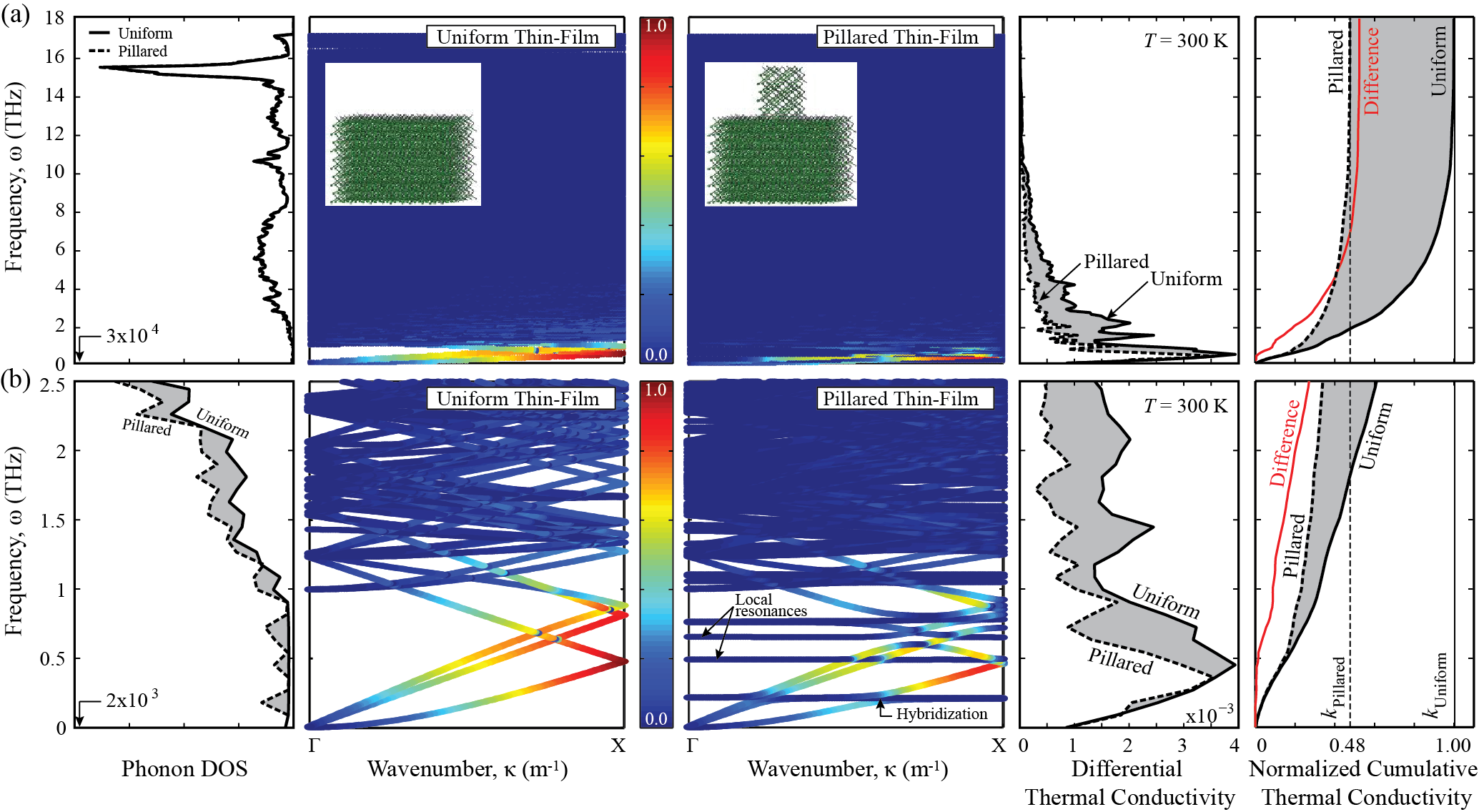}
\caption{Comparison of the phonon dispersion and thermal conductivity of a pillared silicon thin-film with a corresponding uniform thin-film. The dispersion curves are colored to represent the modal contributation to the cumulative thermal conductivity, normalized with respect to the highest modal contribution in either configuration. The full spectrum is shown in (a) and the $0 \leq \omega \leq 2.5$ THz portion is shown in (b). Phonon DOS and the thermal conductivity, in both differential and cumulative forms, are also shown. The grey regions represent the difference in quantity of interest between the two configurations. The introduction of the pillar in the unit cell causes striking changes to all these quantites.}
\label{fig:02}
\end{figure*}
\noindent We obtain the full phonon band structure for a set of suspended uniform silicon thin-films by running atomic-scale lattice dynamics (LD) calculations in which the three-body Tersoff potential~\cite{Tersoff_1988} is used for the Si-Si bonds with only the first nearest neighboring interactions considered. All calculations are conducted after minimizing the interatomic potential energy at constant pressure. For the thermal conductivity predictions, we use the Callaway-Holland (C-H) model~\cite{Callaway_1959,*Holland_1963} which integrates the contribution of each phonon mode along each dispersion branch over a given wavevector path, sums over all available branches and integrates across all other directions to cover the total volume of the supercell's Brillouin zone. Given that a thin-film represents a two-dimensional (2D) material (i.e., phonon wave motion is permitted only in the in-plane direction), the C-H model along the $x$-direction-aligned $\Gamma X$ path takes the form of
\begin{equation}
k = \frac{1}{A_{z} \pi} \sum_{\lambda} \int_{0}^{\pi/A_{x}}C_{ph}( \kappa, \lambda) v_{g}^{2}(\kappa,\lambda)\tau(\kappa,\lambda)\kappa d\kappa
\label{ThC}
\end{equation}
\noindent where $\kappa$, $\lambda$, $C_{ph}$, $v_{g}$ and $\tau$ denote phonon wavenumber, branch index, specific heat, group velocity and scattering time, respectively. The three latter quantities are dependent on the phonon dispersion. The specific heat is expressed as $C_{ph}(\kappa,\lambda) = k_{B} \left(\hbar\omega(\kappa,\lambda)/{k_{B}T}\right)^2f(\kappa,\lambda)$ where $f(\kappa,\lambda)=e^{\hbar\omega(\kappa,\lambda)/k_{B}T}/\left[e^{\hbar\omega(\kappa,\lambda)/k_{B}T}-1\right]^2$, $\omega$ is the frequency, $T$ is the temperature, $k_{B}$ is the Boltzman constant, and $\hbar$ is the adjusted Plank's constant. The group velocity and scattering time parameters represent the underlying mechanisms that determine the rate of phonon transport, that is, respectively, the rate of energy transfer within coherent motion and the time it takes for a nonlinear scattering event to incoherently take place. The former is expressed as $v_{g}(\kappa,\lambda) = \partial\omega(\kappa,\lambda)/{\partial\kappa}$ and the latter as $\tau(\kappa,\lambda)=\left( \tau_{U}(\kappa,\lambda)^{-1} + \tau_{I}(\kappa,\lambda)^{-1} + \tau_{B}(\kappa,\lambda)^{-1}\right)^{-1}$ where $\tau_{U}^{-1}(\kappa,\lambda)=AT\omega^{2}(\kappa,\lambda)e^{-B/T}$, $\tau_{I}^{-1}(\kappa,\lambda)=D\omega^{4}(\kappa,\lambda)$ and $\tau_{B}^{-1}(\kappa,\lambda)=|v_{g}|/L$, representing Umklapp, impurity and boundary scattering, respectively. The parameters $A$, $B$ and $D$ are all obtained empirically. For $A$ and $B$, we utilize measured data for silicon thin-films on a substrate~\cite{Goodson_1999,*Liu_2004,*Liu_2005} since the temperature-dependent trends are similar to their suspended counterparts. Concerning impurity scattering, we use $D=1.32\times 10^{-45} s^{3}$~\cite{Mingo_2003}. The effective boundary scattering length, $L$, is defined as $L=t/(1-p)$ where $p$ is a surface specularity parameter ($0 \leq p \leq 1$)~\cite{Balandin_1998,Mingo_2003}. Due to the high sensitivity of the fitting parameters to the thin-film thickness (especially for very low $t$), we fit our model for a variety of thicknesses, $t = 20, 30,  50,  100,  420$ nm, around a temperature of $T = 300K$. Figure \ref{fig:01}(a) shows the empirical data used (marked with square dots) as well as the results from our model using $p=0$. It is noted that due to the limited availability of data across a wide temperature range, only the solid dots are considered for the fitting to ensure the expected trends at $T = 300K$. Upon obtaining the parameter values for an adequate number of thin-film thicknesses, a second level of curve fitting is performed to harness scattering parameters for a wide range of thin-films as shown in Figs. \ref{fig:01}(b) and \ref{fig:01}(c) (see~\cite{Davis_2012} for further details on this two-step fitting process).\\
\indent We now turn to the main focus of this work which is to determine how the presence of nanoscale resonating pillars reduces the thermal conductivity in a thin-film. For a proof of concept, we first consider a problem where the thin-film thickness is extremely small. The baseline study for this case consists of a thin-film supercell with a square base of $6\times 6$ CCs ($A_{x} = A_{y} = a_{\rm{NPM}} = 3.26$ nm) and a thickness of $M = 5$ CCs ($A_{z} = t = 2.72$ nm); this corresponds to a rectangular solid containing 1440 atoms. The pillar is placed at the top of the thin-film and has a square base of $2\times 2$ CCs (side length of $d = 1.09$ nm) and a height of $3$ CCs ($h = 1.63$ nm) and itself contains $96$ atoms. The geometrical configuration of both supercells is shown in the insets of Fig. \ref{fig:02}(a). The phonon dispersion along the $\Gamma X$ path is presented in the same figure (Fig. \ref{fig:02}(a): full spectrum; Fig. \ref{fig:02}(b): low frequency portion of the spectrum) for both the uniform thin-film and the pillared thin-film. For the thermal conductivity predictions, we keep the Umklapp scattering parameters constant between the uniform and pillared cases. This provides a conservative approximation for the latter since it has been shown that hybridizations cause a slight reduction of phonon lifetimes~\cite{Christensen_NM_2008}. We also keep the boundary scattering parameters constant since the pillars are relatively small in cross-sectional area and are external to the main cross section of the nominal thin-film, and are therefore not expected to significantly deviate from the uniform thin-film boundary scattering parameters. For the present model,we use the appropriate $A$ and $B$ parameters for a $t = 2.72$-nm thin-film as shown in Figs. \ref{fig:01}(b) and \ref{fig:01}(c) and consider the case of $p=0$. The dispersion curves in Fig. \ref{fig:02} are colored in a manner that reflects the contribution to the thermal conductivity of each particular phonon mode (identified by branch index and wavenumber). The colors are normalized so that the participation of the mode with the highest contribution to the thermal conductivity (across both the uniform and pillared thin-film cases) is equal to unity. The phonon frequency density of states (DOS), the differential thermal conductivity, and the normalized cumulative thermal conductivity, all as a function of frequency, are also shown in the figure. \\ 
\indent We note several observations from Fig. \ref{fig:02}: (1) The lower (acoustical) branches contribute to a significant portion of the thermal conductivity in both the uniform and pillared thin-films. In addition, we see that the higher wavenumbers also significantly contribute to the thermal conductivity (contrary to the bulk case). One factor to recall here is that the boundary scattering term has been set to the thin-film thickness, i.e., $L = t = 2.72$ nm. When this value is very small, the long waves (i.e., those near the $\Gamma$-point in the band diagram) are effectively eliminated and hence we get the low contribution at the low wavenumber end of the acoustical branches. (2) The presence of the pillars causes a series of flat locally resonant phonon modes to appear across the entire spectrum, i.e., at both subwavelength and superwavelength frequencies. These modes interact with the underlying acoustic and optical thin-film phonon modes and form a hybridization of the dispersion curves. This in turn leads to a flattening of the branches at the intersections and hence a reduction in the group velocities and the thermal conductivity. The introduction of the pillars reduces the thermal conductivity to $48\%$ of that of the uniform thin-film. This is a remarkable outcome considering that the pillars introduce $288$ new degrees of feedom per unit cell, each of which add one more branch to the summation carried out in Eq.~\eqref{ThC}. Thus even though more heat carriers are added to the system, less energy is actually carried due to the hybridization mechanism. (3) We note that the branches under 1.5 THz (mostly acoustic branches) for the uniform case contribute approximately to $40\%$ of the thermal conductivity. The presence of the pillars significantly modifies the relative contribution of these branches, which now contribute to roughly $60\%$ of the thermal conductivity. With the pillars, nearly $70\%$ of the thermal conductivity is accounted for by phonons below $2.5$ THz compared to $60\%$ without pillars. For the pillared case, the vast majority of this $70\%$ falls within the range $0.5 < \omega \leq 2.5$ THz. The remaining $30\%$ are mostly accounted for in the range $2.5 < \omega \leq 10$ THz. These results indicate that the intense flattening effect caused by the local resonances on the dense high frequency optical modes causes the contribution profile to shift downwards, allowing the acoustic and low frequency optical modes to carry more weight. However, at very high frequencies (above 10 THz), the thin-film dispersion curves are already too flat providing the horizontal resonant branches little opportunity for any noticeable alteration of the group velocities.\\
\indent For the cases considered thus far, we have modeled the dispersion of the thin-film with pillars utilizing atomic-scale LD. However, due to the profound computational intensity associated with solving large complex eigenvalue problems, this type of model is limited to very small sizes (using our current resources, this is roughly in the order of $5$ nm in supercell side length). Given that current nanostructure fabrication technology is practically limited to minimum feature sizes roughly an order of magnitude larger, we turn to a continuum-based finite-element (FE) model for our LD calculations, albeit we pay special attention to the FE resolution in terms of the number of elements per CC, $n_{ele}/\rm{CC}$, when compared to the atomic scale model. To understand the sensitivity of the thermal conductivity prediction to the FE resolution, we include in the Appendix comparisons of results obtained by both FE and atomic-scale LD models. From these results we note that with increased FE resolution, the FE model maintains a consistent trend and approaches the atomic-scale LD model. We also examine in the Appendix the FE performance for a larger model (for which atomic-scale LD results are not available) and again observe a converging trend. \\ 
\begin{figure}[t!]
\centering
\includegraphics[scale= 1]{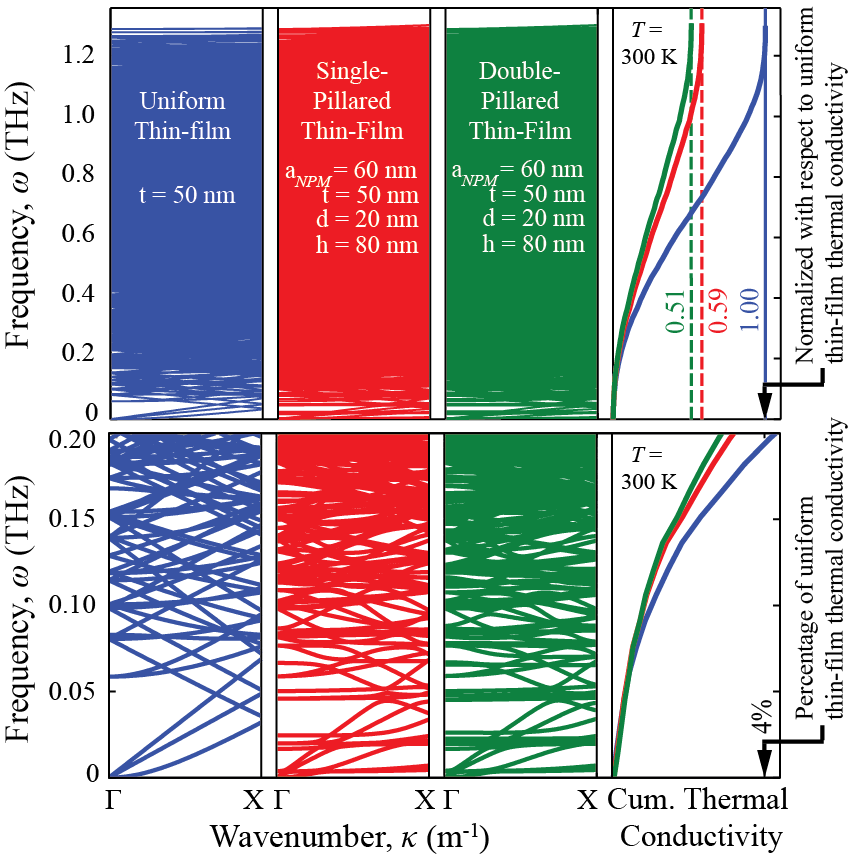}
\caption{Full dispersion comparison of a uniform $t = 50$-nm thin-film (left), with an $80$-nm single pillar (center) and an $80$-nm double pillar (right).  A focus on the first few dispersion branches is also shown as well as the cumulative thermal conductivity as a function of frequency.}
\label{fig:03}
\end{figure}  
\indent Using FE-based lattice dynamics, we now analyze a NPM of a relatively large size: $t = 50$ nm ($A = 6.39\times 10^{-19}$ s/K, $B = 203$ K and $p=0$), $a_{\rm{NPM}} = 60$ nm, $d = 20$ nm and $h = 80$ nm. We set $n_{\kappa} = 129$ and $n_{ele}/\rm{CC} = 0.109$ for this problem. We also analyse a corresponding uniform thin-film for comparison. We again use identical scattering parameters for the pillared and uniform models, noting that this approximation improves with an increase in thin-film thickness. The results appear in Fig. \ref{fig:03} for a NPM with pillars on either one side or on two sides and for the uniform thin-film. We note a few distinctive traits in these results: (1) The maximum frequency remains constant despite the extra branches that get introduced due to the added degrees of freedom of the pillar, and (2) despite this addition of degrees of freedom, once again the NPM has a reduced thermal conductivity ($59\%$ of the uniform thin-film's value) due to the penetration of the local resonance branches into the phonon spectrum. With a higher FE resolution, the predicted reduction is expected to increase, as suggested by Fig. \ref{fig:SM02}. As for the double pillared thin-film, we observe that additional flat branches around the resonant frequencies appear, which causes further reduction in the thermal conductivity, yet the incremental improvement is not as significant as adding pillars on just a single side. We repeat the analysis with a specularity parameter of $p=1$ and obtain $k_{\rm Pillared}/k_{\rm Uniform}$ values of $76\%$ and $73\%$ for single and doubled pillared thin-films, respectively (using recalculated Umklapp scattering parameters, $A = 1.20\times 10^{-18}$ s/K and $B = 15$ K, which were obtained using $p=1$). \\
\indent Locally resonant acoustic metamaterials were introduced by Liu et al.~\cite{liu2000locally} to control acoustic waves. Here we have introduced, for the first time, the concept of a locally resonant NPM to control heat waves. In acoustics, the local resonances couple with the dispersion curves associated with the periodic arrangement of the resonators, or the longwave linear dispersion of the embedding medium when looking only at the subwavelength regime; here the coupling is between the local resonance modes and the atomic-scale dispersion of the underlying crytalline material. Acoustic metamaterials, like their electromagnetic counterparts~\cite{Pendry_IEEE_1999,*Smith_PRL_2000}, derive their unique properties at subwavelength frequencies. In NPMs, the local resonances produce desirable effects across the entire spectrum, including the superwavelength regime. Indeed, we saw that despite the injection of additional phonons (associated with the added degrees of freedom of the resonators), the thermal conductivity has been reduced, and this is attributed to the hybridizations taking place at both subwavelength and superwavelength frequencies. This outcome provides a broader perspective to the definition of a metamaterial. Finally, our proposed NPM configuration$-$which is based on pillared thin-films$-$provides a powerful mechanism for reducing the thermal conductivity without altering the base thin-film material and hence having a minimal effect on the electrical conductivity. This scenario is markedly advantageous for thermoelectric energy conversion. For our models, we obtained a conservative prediction of thermal conductivity reduction by as high as roughly a factor of 2 compared to a corresponding uniform thin-film. Upon analysis with higher resolution models, optimization of dimensions, merging with other thin-film-based thermoelectric materials with proven performances, among other factors, it is perceivable to reach exceedingly high values of $ZT$ using the concept of a nanophononic metamaterial.\\       
\indent The authors would like to thank Dr. E. A. Misawa and the National Science Foundation for support of this research under Grant No. CMMI 0927322 and graduate students L. Liu and O. R. Bilal for their assistance with the FE code programming and mesh set-up, respectively. 


%
%

%


\bibliography{aipsamp5_short}

\renewcommand\thefigure{\thesection A\arabic{figure}} 
\section{Appendix}\label{sec: Appd}
\noindent \bf{Finite-element resolution analysis for thin-film models} \\
\indent \rm We consider a supercell whose thin-film base has a thickness of $A_{z} = t = 3.26$ nm ($A = 4.17\times 10^{-16}$ s/K, $B = 705$ K and $p=0$) and the rest of the dimensions as given in Fig. \ref{fig:SM01}. These dimensions are selected to enable a comparison with an atomic-scale LD supercell model with a thin-film base composed of $6$ CC and a pillar base and height formed from $2$ and $4$ CC, respectively. For the FE model, we use three-dimensional cubic elements. In Fig. \ref{fig:SM01}, we directly compare the reduction in the thermal conductivity for the thin-film with pillars normalized with respect to the uniform case, for various FE resolutions, $n_{ele}/\rm{CC}$, and wavenumber discretization resolutions, $n_{\kappa}$. First, we find that as we increase $n_{\kappa}$ (which numerically improves the prediction of the C-H model), the thermal conductivity converges to a constant value. Second, when $n_{ele}/\rm{CC}$ is increased, the reduction in the thermal conductivity due to the presence of the pillars increases and also converges to a constant value. Finally, we note that with increased FE resolution, the FE model maintains a consistent trend and approaches the atomic-scale LD model. This provides confidence that, for a given resolution, the FE model conservatively captures the nanoscale phonon dynamics behavior as far as the effects of the pillars on the overall dispersion, and hence the thermal conductivity reduction, are concerned.\\
\indent Upon proceeding to a thin-film model with a larger thickness, it is difficult to maintain the same level of FE resolution due to limitation of computational resources. To examine the convergence peformance under such limitation, we now analyse a larger thin-film model using substantially lower $n_{ele}/\rm{CC}$ values. We choose uniform and pillared thin-films with a thickness of $t = 60$ nm ($A = 5.90\times 10^{-19}$ s/K, $B = 200$ K and $p=0$). The NPM supercell here has a base length of $a_{\rm{NPM}} = 60$ nm, a pillar width of $d = 20$ nm and a pillar height of $h = 40$ nm. The results are shown in Fig. \ref{fig:SM02} where we observe that the reduction in the thermal conductivity also displays a converging trend as that is shown in Fig. \ref{fig:SM01} except that the rate of convergence is slower. This in fact suggests that if the $n_{ele}/\rm{CC}$ resolution is increased further, a substantial additional reduction in the thermal conductivity of the NPM compared to the uniform case is to be expected.\\ 
\setcounter{figure}{0}  
\begin{figure}[H]
\centering
\includegraphics[scale= 1]{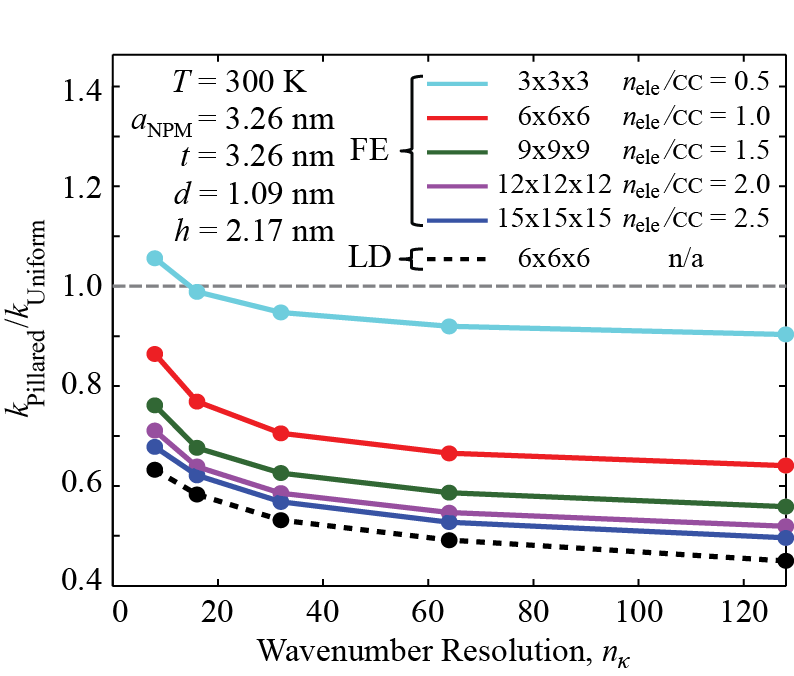}
\caption{Comparison of the thermal conductivity of a $t=3.26$-nm thin-film with and without the presence of pillars utilizing FE of varying $n_{ele}/\rm{CC}$ and $n_{\kappa}$ resolutions (solid lines). The unit cell dimensions of the FE model are equivalent to a corresponding atomic-scale LD model (dashed line) to enable a direct comparison.}
\label{fig:SM01}
\end{figure} 
\begin{figure}[H]
\centering
\includegraphics[scale= 1]{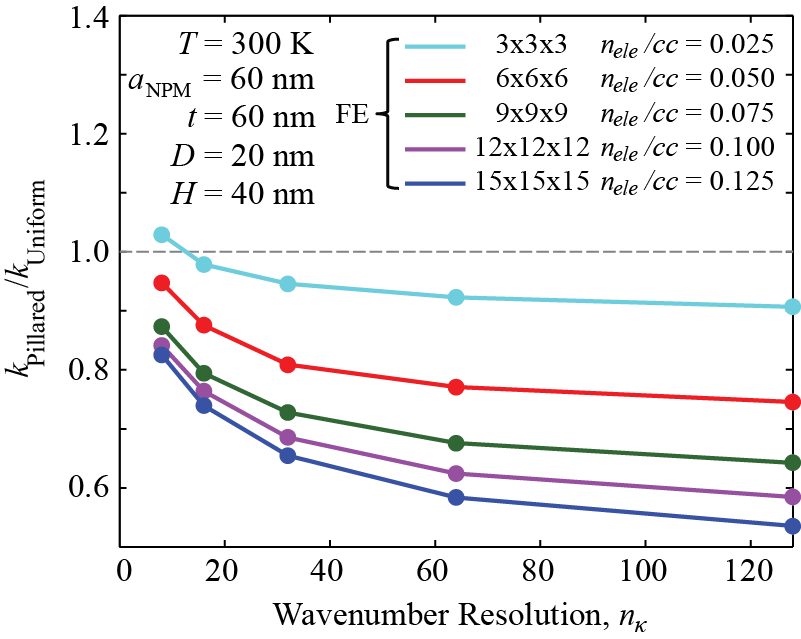}
\caption{Comparison of the thermal conductivity of a $60$-nm thin-film with and without the presence of pillars utilizing FE of varying $n_{ele}/\rm{CC}$ and $n_{\kappa}$ resolutions. The steady but slow rate of convergence observed indicates that upon further increase in FE resolution, the relative thermal conductivity, $k_{\rm Pillared}/k_{\rm Uniform}$, is expected to decrease substantially.}
\label{fig:SM02}
\end{figure}    


\end{document}